\documentclass[aps,prb,twocolumn,footinbib,groupedaddress]{revtex4-1} 
	
	\usepackage{amsmath}
	\usepackage{xcolor}
	\usepackage{amssymb}
	\usepackage{graphicx} 
	\usepackage[breaklinks=true,colorlinks,citecolor=blue,linkcolor=blue,urlcolor=blue]{hyperref}
	\usepackage{physics}
	\usepackage{bbold}
	\usepackage{soul}
   \usepackage{enumerate}
    \usepackage{color}
    \usepackage{ulem}

\begin{document}
\title{Thermodynamic uncertainty relations for systems with broken time reversal symmetry: The case of superconducting hybrid systems}

\author{Fabio Taddei}
\affiliation{NEST, Istituto Nanoscienze-CNR and Scuola Normale Superiore, I-56126 Pisa, Italy}

\author{Rosario Fazio}
\affiliation{The Abdus Salam International Center for Theoretical Physics,  34151 Trieste, Italy}
\affiliation{Dipartimento di Fisica, Universit\`a di Napoli ``Federico II'', I-80126 Napoli, Italy}

\begin{abstract}
We derive new bounds to the thermodynamic uncertainty relations (TURs) in the linear-response regime for steady-state transport in two-terminal systems when time reversal symmetry (TRS) is broken.
We find that such bounds are different for charge and heat currents and depend on the details of the system, through the Onsager coefficients, and on the ratio between applied voltage and temperature difference.
As a function of such a ratio, the bounds can take any positive values.
The bounds are then calculated for a hybrid coherent superconducting system using the scattering approach, and the concrete case of an Andreev interferometer is explored.
Interestingly, we find that the bound on the charge current is always smaller than 2 when the system operates as a heat engine, while the bound on the heat current is always larger than 2 when the system operates as a refrigerator.
\end{abstract}

\maketitle

\section{Introduction}
Constraints on the performance of thermodynamic machines are set by thermodynamics laws. 
The most important example is the bound on the efficiency of a thermal machine exchanging heat between two reservoirs, which is represented by the Carnot efficiency and based on the second law.
In the context of chemical reactions, Barato and Seifert, in Ref.~\onlinecite{Barato2015}, introduced a new constraint which
sets a tradeoff between the minimal relative uncertainty of the output of a reaction and the minimal energy cost to generate such an output.
This has been expressed through the inequality
\begin{equation}
\label{tur1}
Q^I=\frac{S_I\sigma}{I^2k_B}\geq Q_{\rm bound}^I,
\end{equation}
where $I$ is the current, $S_I$ is the current noise, $\sigma$ is the entropy production rate and $k_B$ is the Boltzmann constant.
Such an inequality was termed thermodynamic uncertainty relation (TUR) and assesses the minimal dissipation (measured by $\sigma$) required to generate an output current ($I$) with minimum relative uncertainty.
In Ref.~\onlinecite{Barato2015}, using biased random walks in the stationary state, it has been proven that $Q_{\rm bound}^I=2$ within the linear response theory.
Such a bound has been also derived for Markov jump processes and networks~\cite{Gingrich2016,Pietzonka2016,Polettini2016} (even beyond linear response).
Since then a large numbers of papers have addressed TURs in various systems and approaches, even proposing more general inequalities~\cite{Timpanaro2019,Hasegawa2019,Guarnieri2019,Dechant2019}.
Experimental evidence of the bound $Q_{\rm bound}^I=2$ has been reported for atomic-scale quantum conductors~\cite{Friedman2020}, while the possibility of its violation has been reported in Ref.~\onlinecite{Pal2020} for a two-qubit system using a NMR setup.
However, there are situations where the value of the bound is not $Q_{\rm bound}^I=2$.
One example is when time reversal symmetry (TRS) is broken.
In this case, various inequalities for quantities related to $Q^I$ have been derived, see Refs.~\onlinecite{Brandner2018,Macieszczak2018,Proesmans2019,Chun2019,Cangemi2021,Park2021,Saryal2022}.
\begin{figure}[!tbh]
	\centering
	\includegraphics[width=0.9\columnwidth]{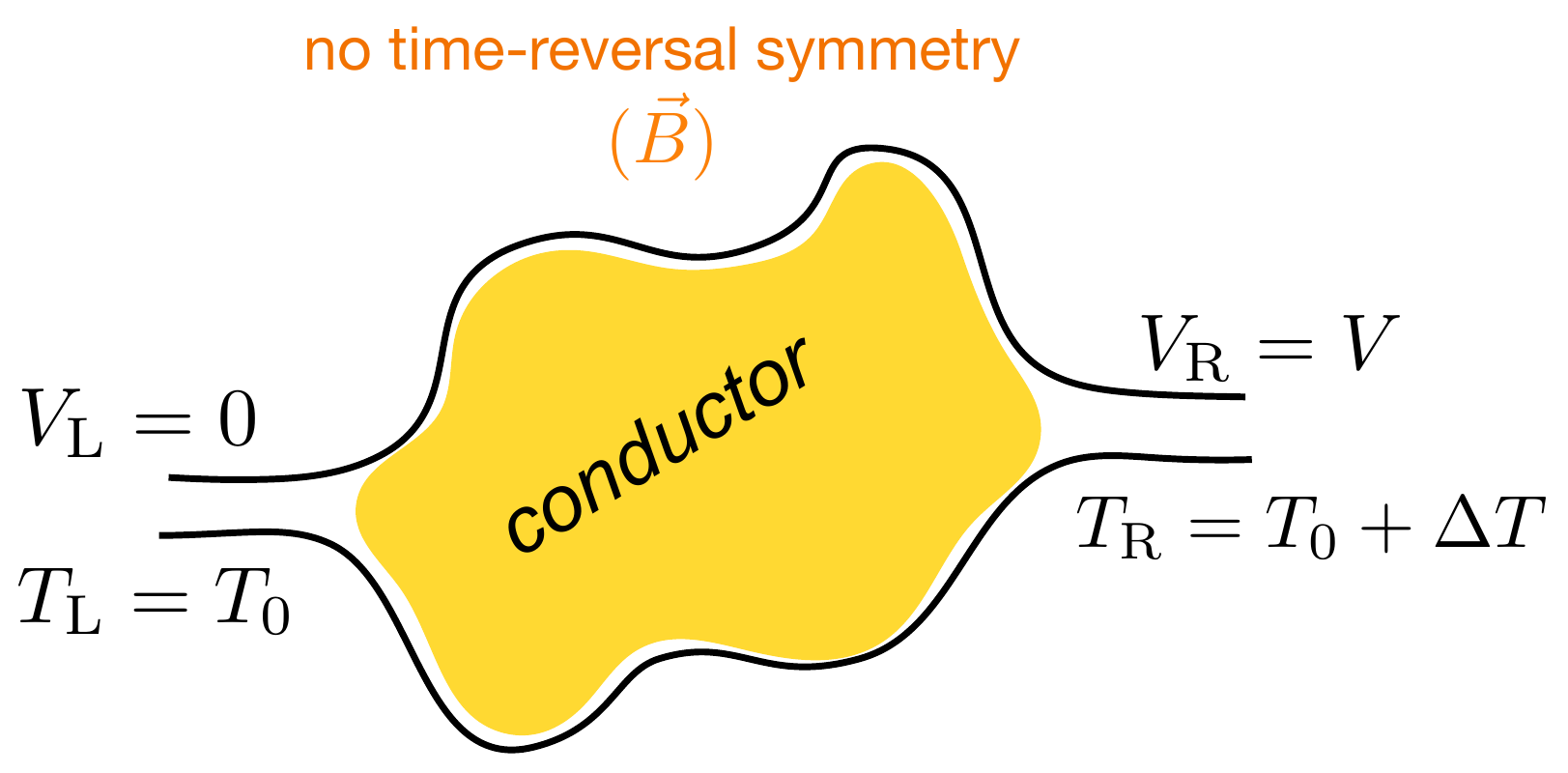}
	\caption{Sketch of a two-terminal system with broken TRS, where a voltage $V$ and a temperature difference $\Delta T$ are applied between the right and the left electrode.}
	\label{first}
\end{figure}

Here we are interested in the case where TUR is formulated for electric and thermal transport in the steady-state regime, as in Refs.~\onlinecite{Brandner2018,Agarwalla2018,Ptaszynski2018,Macieszczak2018,Saryal2019,Proesmans2019,Guarnieri2019,Kheradsoud2019,Liu2019,Friedman2020,Saryal2021,Ehrlich2021,Gerry2022,Saryal2022}.
In particular, we consider a generic two-terminal system, possibly containing superconducting regions, as depicted in Fig.~\ref{first}.
The right electrode is kept at voltage $V_{\rm R}=V$ and temperature $T_{\rm R}=T_0+\Delta T$, while the left electrode is grounded ($V_{\rm L}=0$) and kept at the reference temperature $T_{\rm L}=T_0$.
Charge current ($I$) and heat current ($J$) flowing in the right electrode are expressed, in the linear-response regime, through the Onsager coefficients as
\begin{align}
\label{Ons}
I=L_{11}\frac{V}{T_0}+L_{12}\frac{\Delta T}{T_0^2}  \\ 
J=L_{21}\frac{V}{T_0}+L_{22}\frac{\Delta T}{T_0^2} , \nonumber 
\end{align}
where $L_{ij}$ are the components of the Onsager matrix.

Remarkably, these systems can operate as thermoelectric steady-state thermal machines~\cite{Benenti2017}, such as heat engines or refrigerators.
On the one hand, a heat engine is a machine which converts heat into work.
This happens when a temperature difference applied to the conductor gives rise to a heat current flowing out of the hot electrode and to an electric current flowing against the applied voltage, thus generating a usable electrical power.
The efficiency of a heat engine is defined as the ratio between the output power and the absorbed heat. 
On the other hand, a refrigerator is a thermal machine which uses external work to extract heat from a cold thermal bath.
It is obtained when a  conductor is subjected to a voltage which gives rise to a heat current that flows against the temperature difference, i.e.~extracting heat from the cold electrode.
Interestingly, in Ref.~\onlinecite{Pietzonka2018} is was shown that the TUR can be written as a bound on the efficiency of a heat engine.

For quantum thermoelectric devices in Refs.~\onlinecite{Ehrlich2021,Timpanaro2021,Gerry2022} it was shown, within the scattering theory, that the inequality (\ref{tur1}) can be violated, far from linear response, at an arbitrary extent when the transmission function features sharp steps (as boxcar shapes).
The largest violations occur for idealized transmission functions consisting of a collection of rectangular functions~\cite{Timpanaro2021}.
On the other hand, as shown in Refs.~\onlinecite{Kheradsoud2019,Gerry2022}, no violations of the inequality (\ref{tur1}) occur when the transmission functions is smooth.
This can be the case, for example, for systems such as quantum dots-based devices~\cite{Gerry2022} and quantum points contacts~\cite{Kheradsoud2019}.
The case of systems where TRS is broken has been considered in Refs.~\onlinecite{Brandner2018,Macieszczak2018,Proesmans2019,Saryal2022}.
Different forms of TURs (for quantities related to $Q^I$) have been derived using different criteria (though mainly based on the non-negativity of the net entropy production rate).

\section{Summary of the results}
\label{summ}
In this paper, for two-terminal systems in the absence of TRS, we derive new bounds for the quantity $Q^I$, defined in Eq.~(\ref{tur1}), and for the quantity
\begin{equation}
\label{tur5}
Q^J=\frac{S_J\sigma}{J^2k_B} ,
\end{equation}
where $J$ is the heat current and $S_J$ is the zero-frequency heat current noise.
Namely, within the the linear-response regime, we obtain 
\begin{align}
\label{turr}
Q_{\rm bound}^I= 2
\left[ 1+\frac{1}{2}\frac{L_{21}-L_{12}}{L_{12}+L_{11}\left ( \frac{VT_0}{\Delta T} \right)}
\right]^2 
\end{align}
and
\begin{align}
\label{turrh}
Q_{\rm bound}^J= 2
\left[ 1+\frac{1}{2}\frac{L_{12}-L_{21}}{L_{21}+L_{22}\left ( \frac{\Delta T}{V T_0} \right)}
\right]^2 ,
\end{align}
for charge and heat current, respectively.
Here $L_{ij}$ are the components of the Onsager matrix, while $V$ and $\Delta T$ are the voltage and temperature biases applied to the conductor depicted in Fig.~\ref{first}.
It should be stressed that $L_{12}$ and $L_{21}$ are different in the absence of TRS.
Remarkably, the two expressions (\ref{turr}) and (\ref{turrh}) depend explicitly on the ratio between the biases, are different for charge and heat currents, and reduce to the usual bound $Q_{\rm bound}^I=Q_{\rm bound}^J= 2$ when TRS is restored, i.e.~when $L_{12}=L_{21}$.
In particular, depending on properties of the system (encoded in the Onsager matrix) and on the biases, the two bounds can be arbitrarily larger or smaller than 2, i.e.~tighter or looser than in the TR symmetric case.
\begin{figure}[!tbh]
	\centering
	\includegraphics[width=1.0\columnwidth]{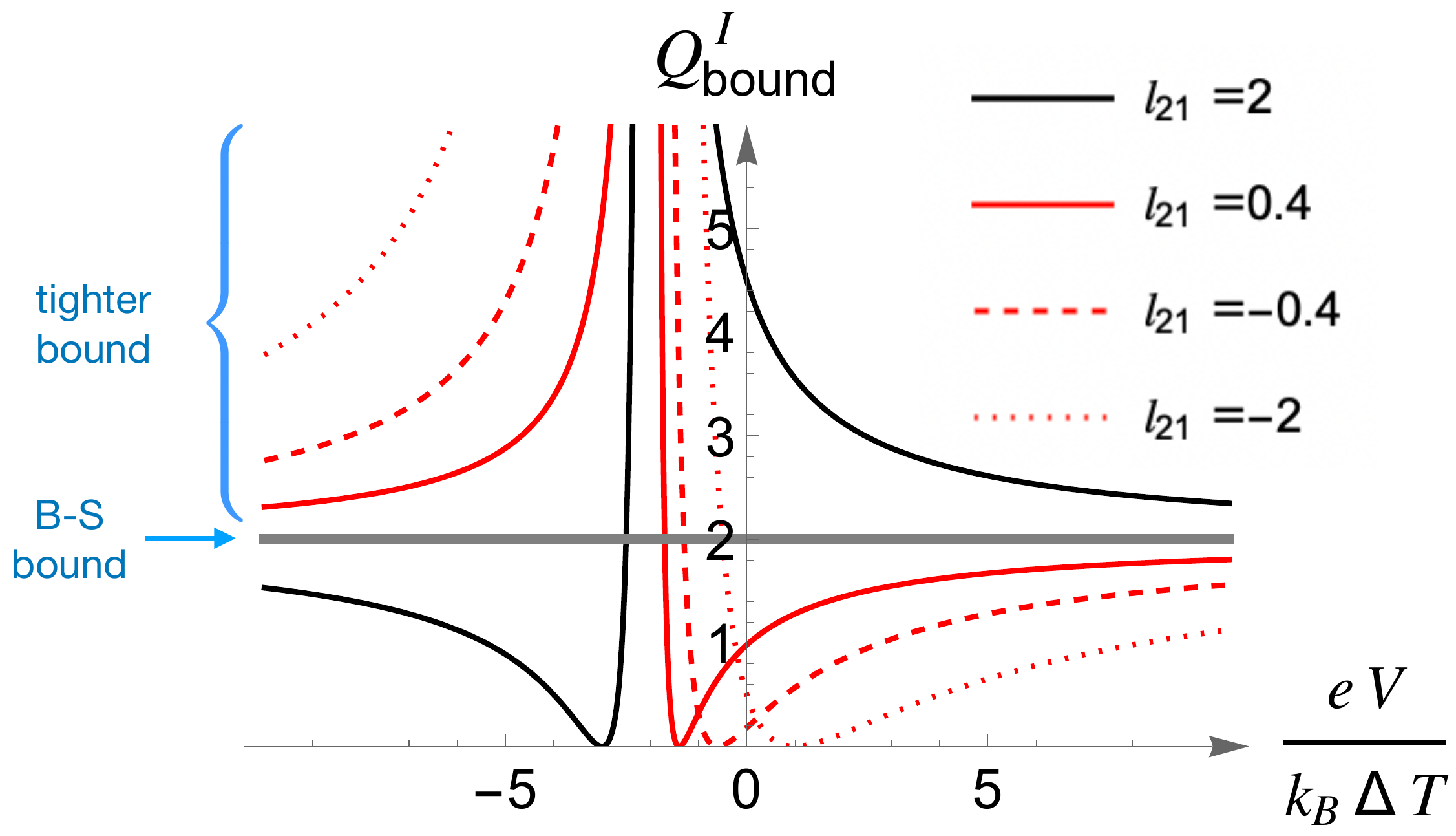}
	\caption{Plots of  $Q^I_{\rm bound}$ as a function of the dimensionless ratio $eV/(k_B\Delta T)$ for different parameters settings: $l_{21}=2, 0.4,-0.4,-2$ for solid black, solid red, dashed red and dotted red curve, respectively, and $l_{11}=0.5 e/(k_BT_0)$. The thick gray horizontal line is the reference Barato-Seifert (B-S)~\cite{Barato2015} value $Q^I_{\rm bound}=2$ occurring for $l_{21}=1$, which is the TR symmetric case. Above (below) the value 2 the bound is tighter (looser) than the B-S bound.}
	\label{second}
\end{figure}

Before proceeding, we illustrate the behavior of the new bounds by considering different choices of Onsager matrices, i.e.~different systems.
Here we concentrate on $Q_{\rm bound}^I$.
First notice that dividing the second term in Eq.~(\ref{turr}) by $L_{12}$, the bound depends on two parameters only, namely $L_{11}/L_{12}\equiv l_{11}$ and $L_{21}/L_{12}\equiv l_{21}$, so that
\begin{align}
\label{turr2}
Q_{\rm bound}^I= 2
\left[ 1+\frac{1}{2}\frac{l_{21}-1}{1+r l_{11}}
\right]^2 ,
\end{align}
and $r= VT_0/\Delta T$.
On the one hand, $l_{11}$ simply acts as a normalization factor for $r$ and, in particular, yields the transformation $Q_{\rm bound}^I(r)\rightarrow Q_{\rm bound}^I(-r)$ when reversing the sign of $l_{11}$.
On the other hand, $l_{21}$ determines the zeros of the bound.
Therefore, the behavior of the bound can be understood by considering a few values of $l_{21}$ and a fixed value of $l_{11}$. 
As we shall see below, the two parameters can take any values, both positive and negative.
In Fig.~\ref{second} we plot $Q_{\rm bound}^I$ as a function of the dimensionless ratio $\tilde{r}=eV/(k_B\Delta T)$.
The different curves are relative to $l_{21}=2, 0.4,-0.4,-2$ (black, red, dashed red and dotted red curve, respectively) and $l_{11}=0.5 e/(k_BT_0)$.
The gray thick horizontal line is the reference, Barato-Seifert (B-S)~\cite{Barato2015} , value $Q_{\rm bound}^I=2$ obtained for $l_{21}=1$, corresponding to the TR symmetric case.
First, we notice that all curves diverge for the same value of $\tilde{r}=\tilde{r}_{\infty}\equiv -e/{l_{11}k_BT_0}$, and present a minimum where $Q_{\rm bound}^I=0$.
The black curve, which is the only one for which $l_{21}>1$, is non-monotonous for $\tilde{r}<\tilde{r}_{\infty}$ (where mainly remains smaller than 2), and decreases monotonously for $\tilde{r}>\tilde{r}_{\infty}$ remaining larger than 2.
On the contrary, all red curves, for which  $l_{21}<1$, have an opposite behavior being monotonously increasing for $\tilde{r}<\tilde{r}_{\infty}$, and non-monotonous for $\tilde{r}>\tilde{r}_{\infty}$.
The red curves differ only by the position of the zero and their smoothness.

It is now interesting to notice that, for a coherent quantum conductor with two electrodes, the scattering theory imposes that $L_{12}=L_{21}$ even when TRS is broken~\cite{Benenti2017}.
This implies that the two bounds reduce to the B-S one, namely $Q_{\rm bound}^I=Q_{\rm bound}^J= 2$.
There are, however, two-terminal coherent systems where $L_{12}$ and $L_{21}$ can be different: the ones containing superconducting regions.
(We note in passing that TUR has hardly been studied for hybrid superconducting systems~\cite{Lopez2023,Manzano2023}).
In order to analyze the behavior of the bounds in this case we consider an Andreev interferometer (see Fig.~\ref{AI}), which is the simplest hybrid superconducting system where TRS is broken (by a superconducting phase difference).
In order to calculate the Onsager matrix for such a system we use the scattering theory~\cite{Lambert1998}.
Interestingly, we find that the bound on $Q^I$ is always smaller than 2 when the Andreev interferometer operates as a heat engine, while the bound on $Q^J$ is always larger than 2 when the Andreev interferometer operates as a refrigerator.
In general, it is worth emphasizing that the absence of TRS in a conductor can be detected by looking for values of $Q^I$ or $Q^J$ smaller than 2 in the linear response regime, because they violate the bound for a TRS system.

The paper is organized as follows.
In Sec.~\ref{SecDer} we sketch the derivation of the inequalities expressing the TURs and we examine their properties, while in Sec.~\ref{SecMod} we detail the scattering approach used for describing a two-terminal hybrid superconducting system. Finally, in Sec.~\ref{SecAndInt} we apply the results of the previous sections to an Andreev interferometer by calculating its Onsager coefficients and bounds of the TURs, and we draw the conclusions in Sec.~\ref{SecConc}.

\section{Derivation of the bounds}
\label{SecDer}
 We consider the two-terminal system depicted in Fig.~\ref{first}, possibly containing superconducting regions, in the linear-response regime.
We do not enforce TRS, while we allow for the presence of an applied magnetic field $\mathbf{B}$.
In such a case we recall that, in general, one has $L_{12}\ne L_{21}$, since the Onsager relation actually requires $L_{12}(\mathbf{B})=L_{21}(-\mathbf{B})$.
On the other hand, the entropy production rate in the linear response can be written as
\begin{align}
\label{sig}
\sigma&=IX_\mu+JX_T\nonumber\\
&=L_{11}X_\mu^2+L_{22}X_T^2+( L_{12}+L_{21})X_\mu X_T,
\end{align}
where $X_\mu=V/T_0$ and $X_T=\Delta T/T_0^2$ are the affinities.

We now express the quantity $Q^I$ defined in Eq.~(\ref{tur1}) in terms of the Onsager coefficients by replacing the entropy production rate $\sigma$ and $I$ using Eqs.~(\ref{sig}) and (\ref{Ons}), respectively, obtaining
\begin{align}
\label{Qapprox}
Q^I\simeq 2L_{11}\frac{L_{11}V^2T_0^2+L_{22}(\Delta T)^2+(L_{12}+L_{21})V\Delta T T_0}{(L_{11}VT_0+L_{12} \Delta T)^2} .
\end{align}
Here we have used the fact that, in the linear response (at equilibrium), $S_I\simeq 2k_BL_{11}$  as a consequence of the fluctuation-dissipation relations.
We then express $L_{22}$ in terms of $Q^I$ and the other quantities and, recalling that $L_{11}\geq 0$, we impose the fact that the Onsager matrix is positive semi-definite
\begin{equation}
\label{spOns}
L_{22}\geq \frac{(L_{12}+L_{21})^2}{4L_{11}}.
\end{equation}
From such an inequality, which follows from the positivity of entropy production and does not assume $L_{12}=L_{21}$, one finds Eq.~(\ref{turr}).
Since a small value of $Q^I$ is desirable, one should look for systems where $L_{12}\ne L_{21}$ in such a way that the bound is well below 2.

Similarly, we can derive a bound for $Q^J$, defined in Eq.~(\ref{tur5}), relative to the heat current.
Again, by using Eqs.~(\ref{Ons}) and (\ref{sig}) and imposing the same inequality (\ref{spOns}), but written for $L_{11}$
\begin{align}
L_{11}\geq\frac{(L_{12}+L_{21})^2}{4L_{22}},
\end{align}
(recall that $L_{22}\geq 0$) we obtain Eq.~(\ref{turrh}).
Here we have used the equilibrium relation $S_J\simeq 2k_BL_{22}$.
As far as we know, the two inequalities (\ref{turr}) and (\ref{turrh}) have not been reported so far.

A few comments are in order.
First, note that the two bounds (\ref{turr}) and (\ref{turrh}) are different, i.e. charge and heat currents obey different TURs.
This is not the case for the inequalities derived for TRS breaking systems, for example, in Ref.~\onlinecite{Macieszczak2018} and in Ref.~\onlinecite{Proesmans2019}.
Second, we have an explicit dependence on the affinities through their ratio $V/\Delta T$.
In particular, when $\Delta T=0$, Eq.~(\ref{turr}) reduces to $Q_{\rm bound}^I= 2$, while Eq.~(\ref{turrh}) becomes
\begin{align}
Q^J_{\rm bound}= 2\left( \frac{L_{12}+L_{21}}{2L_{21}} \right)^2.
\end{align}
On the other hand, when $V=0$ Eq.~(\ref{turr}) becomes
\begin{align}
Q_{\rm bound}^I= 2\left( \frac{L_{12}+L_{21}}{2L_{12}} \right)^2 ,
\end{align}
while Eq.~(\ref{turrh}) reduces to $Q^J_{\rm bound}= 2$.

Finally, both bounds (\ref{turr}) and (\ref{turrh}) reduce to the B-S one, namely  $Q_{\rm bound}^I=Q_{\rm bound}^J= 2$, when $L_{12}=L_{21}$.
This is the case for systems with TRS, where such equality constitutes the Onsager reciprocal relation~\cite{Onsager1931}.
Interestingly, the equality $L_{12}=L_{21}$ holds also for two-terminal systems which break TRS, but in which transport is coherent and can be described through the scattering theory~\cite{Benenti2017}, see App.~\ref{AppNormal}.
Thus, the bounds (\ref{turr}) and (\ref{turrh}) reduce to the B-S one for any coherent two-terminal conductor.

There is, however, a notable exception constituted by hybrid superconducting systems.
In this case, as we shall see in the next section, $L_{12}$ can be different from $L_{21}$ when TRS is broken.

\section{Superconducting multi-channel systems}
\label{SecMod}
Let us now consider a hybrid system consisting of a conductor, possibly containing superconducting regions, attached to a normal (N) electrode on the right and a normal or superconducting (S) electrode on the left, see Fig.~\ref{first}.
We use the scattering approach~\cite{Claughton1996,Lambert1998} to calculate the charge and heat currents produced by a voltage and a temperature difference and flowing in the N right lead, see App.~\ref{AppSuper}.
In the linear-response regime, where $eV\ll k_BT_0$ and $\Delta T\ll T_0$, the Onsager coefficients can be expressed as
\begin{widetext}
\begin{align}
\label{L11}
L_{11}=\frac{e^2T_0}{h} \int_{0}^{\infty}dE\; [N^+(E)-R^{++}(E)+R_a^{+-}(E)+N^-(E)-R^{--}(E)+R_a^{-+}(E)] \left( -\frac{\partial f}{\partial E}\right),
\end{align}
\begin{align}
\label{L22}
L_{22}=\frac{T_0}{h} \int_{0}^{\infty}dE\; E^2  [N^+(E)-R^{++}(E)-R_a^{+-}(E)+N^-(E)-R^{--}(E)-R_a^{-+}(E)] \left( -\frac{\partial f}{\partial E}\right),
\end{align}
\begin{align}
\label{L12}
L_{12}=\frac{eT_0}{h} \int_{0}^{\infty}dE\; E  \left[N^+(E)-R^{++}(E)-N^-(E)+R^{--}(E)- R_a^{+-}(E)+R_a^{-+}(E)\right] \left( -\frac{\partial f}{\partial E}\right) 
\end{align}
and
\begin{align}
\label{L21}
L_{21}=\frac{eT_0}{h} \int_{0}^{\infty}dE\; E [N^+(E)-R^{++}(E)-N^-(E)+R^{--}(E)+R_a^{+-}(E)-R_a^{-+}(E)] \left( -\frac{\partial f}{\partial E}\right) .
\end{align}
\end{widetext}
Here 
$N^\alpha(E)$ is the number of open channels, in the N right lead, at energy $E$ for a $\alpha$-like quasiparticle ($\alpha=+$ stands for electron and $\alpha=-$ stands for hole) and $f(E)$ is the Fermi distribution function at temperature $T_0$ and zero chemical potential.
Moreover, $R^{\alpha\alpha}(E)$ is the total energy-dependent normal reflection coefficient for a $\alpha$-like quasiparticle (obtained summing up the probabilities of all open channels), while $R_a^{\alpha\beta}(E)$ is the total Andreev reflection coefficient for a $\beta$-like quasiparticle to be reflected into a $\alpha$-like quasiparticle.
According to Eqs.~(\ref{L12}) and (\ref{L21}), $L_{12}\ne L_{21}$ as long as $R_a^{+-}(E)\ne R_a^{-+}(E)$, i.e. when TRS is broken~\cite{Lambert1998},  and the energy dependence of $R_a^{+-}(E)$ is strong enough.
As we shall see below, these conditions can be met in an Andreev interferometer~\cite{Claughton1996} with two resonant barriers.
On the contrary, when TRS is preserved we have $R_a^{+-}(E)=R_a^{-+}(E)$, or equivalently $R_a^{+-}(E)=R_a^{+-}(-E)$, thus implying that $L_{12}=L_{21}$~\cite{nota}.
It is worthwhile expressing the difference $L_{21}-L_{12}$ as
\begin{align}
L_{21}-L_{12}=
\frac{2eT_0}{h} \int_{-\infty}^{\infty}dE\; E\;R_a^{+-}(E)  \left( -\frac{\partial f}{\partial E}\right) .
\end{align}

Two peculiar cases, which are possible in a hybrid system, are worth mentioning.
First, in the absence of normal reflection, i.e.~when $R^{\alpha\alpha}=0$, we have $L_{12}=-L_{21}$ so that the bound (\ref{turr}) becomes
\begin{align}
Q_{\rm bound}^I= 2
\left[ 1+\frac{L_{12}}{L_{11}\left ( \frac{VT_0}{\Delta T} \right)}
\right]^{-2} .
\end{align}
Second, in the case where the left lead is superconducting and at low temperature ($k_BT_0\ll\Delta$) one finds $L_{12}=L_{21}=L_{22}\simeq 0$, see App.~\ref{AppSuper}, yielding $Q^I\simeq 2$ from Eq.~(\ref{Qapprox}).

In the next section we will use the expressions of the Onsager coefficients, Eqs.~(\ref{L11})-(\ref{L21}), to calculate for a specific systems the bounds in Eqs.~(\ref{turr}) and (\ref{turrh}).

\section{Andreev interferometer}
\label{SecAndInt}
\begin{figure}[!tbh]
	\centering
	\includegraphics[width=0.9\columnwidth]{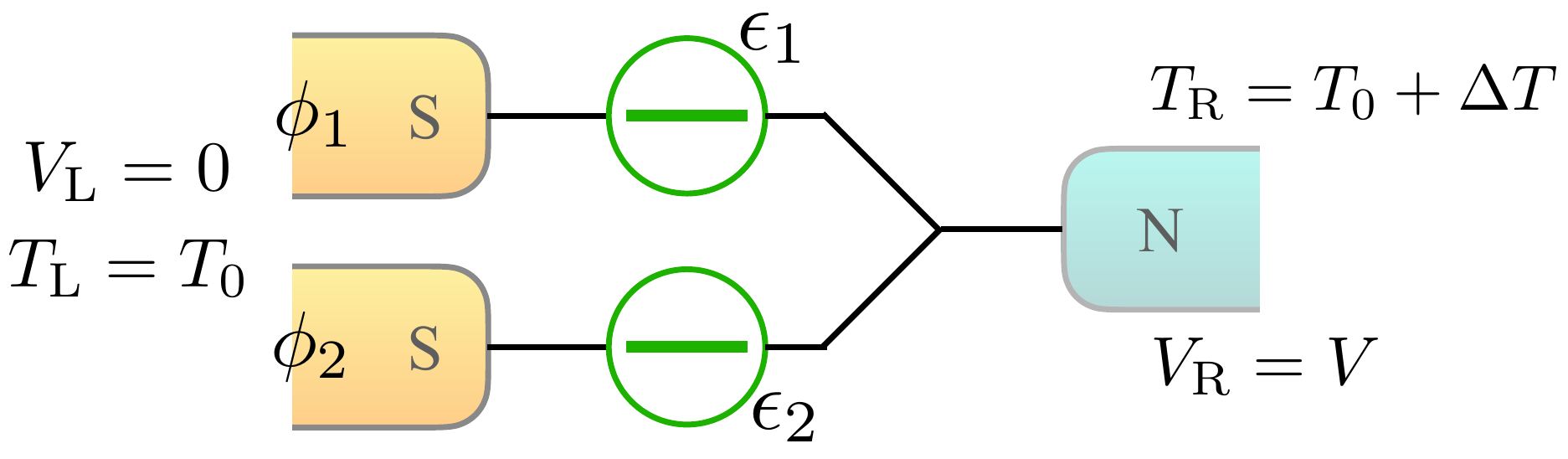}
	\caption{Sketch of an Andreev interferometer, which is composed of a three-leg beam splitter and two resonant barriers (such as quantum dots). The latter are placed on the two parallel left legs, which are connected to the two superconducting (S) regions with different phases ($\phi_1$ and $\phi_2$), while the right leg is connected to a normal (N) electrode. The S electrode is grounded and at the reference temperature $T_0$, while the N electrode is kept at a voltage $V$ and temperature $T_0+\Delta T$.}
	\label{AI}
\end{figure}
Let us now consider an Andreev interferometer, see Fig.~\ref{AI}, consisting of a three-leg beam splitter with two resonant barriers (in green) in the branches on the left connected to two superconducting (S, in yellow) regions characterized by different superconducting phases $\phi_1$ and $\phi_2$ and equal gap $\Delta$.
TRS is broken when $\phi_1\ne\phi_2$.
Note that the system corresponds to a two-terminal setup because the S regions are kept at the same temperature $T_0$ and are grounded (zero chemical potential).
The normal (N) terminal on the right, instead, is kept at a finite voltage $V$ and at a temperature $T_0+\Delta T$.

\begin{figure}[!htb]
	\centering
	\includegraphics[width=0.9\columnwidth]{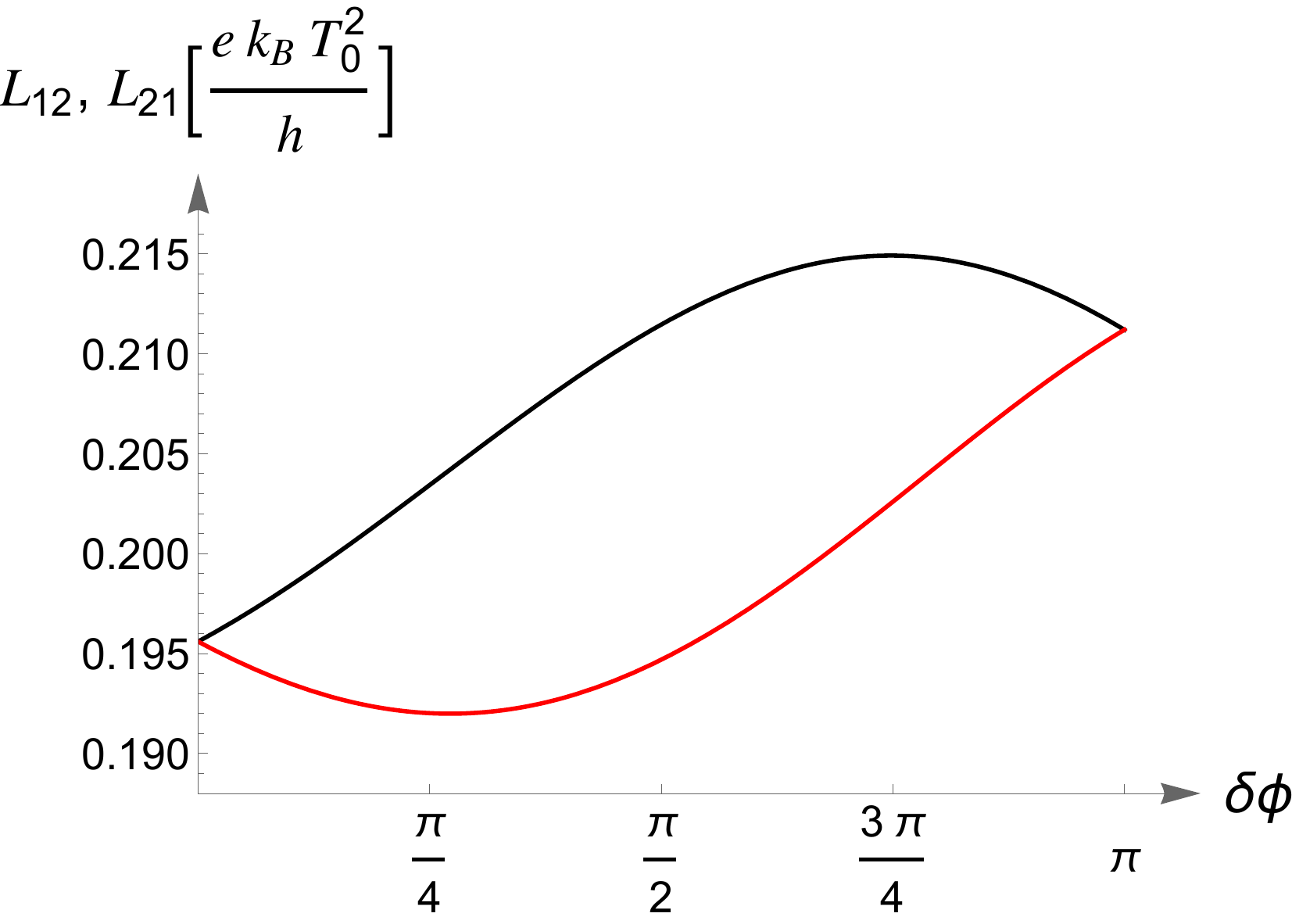}
	\caption{The two Onsager coefficients $L_{12}$ (black) and $L_{21}$ (red) are plotted in units of $ek_BT_0^2/h$ as functions of $\delta\phi$, for $\gamma_1=0.5\Delta$, $\gamma_2=1.0\Delta$, $\epsilon_1=0.4\Delta$, $\epsilon_2=1.9\Delta$, $\Gamma=0.3$, $s_1=1$, $s_2=-1$ and $k_BT_0=0.7\Delta$.}
	\label{phi}
\end{figure}

The scattering amplitudes of the system are calculated by composing the scattering amplitudes of the individual components, i.e.~a three-leg beam splitter, two resonant barriers and two NS perfect interfaces (see App.~\ref{AppScattering} for details).
We denote by $\epsilon_1$ and $\epsilon_2$ the energies of the two resonances and by $\gamma_1$ and $\gamma_2$ their broadenings.
The three-leg beamsplitter is characterized by the transmission probability $\Gamma$ and two constants $s_1,s_2=\pm 1$.
When the energy $E<\Delta$, the scattering matrix contains only normal reflections and Andreev reflections and unitarity imposes that $R_a^{+-}(E)=R_a^{-+}(E)$.
However, provided that $\phi_1\ne\phi_2$, one finds that $R_a^{+-}(E)\ne R_a^{-+}(E)$ for $E>\Delta$ when $\epsilon_1\ne\epsilon_2$ or $\gamma_1\ne\gamma_2$, implying that $L_{12}\ne L_{21}$.
In general the difference $L_{12}-L_{21}$ has no definite sign.
As an example, in Fig.~\ref{phi}
we plot $L_{12}$ (black) and $L_{21}$ (red) as functions of $\delta\phi=\phi_1-\phi_2$
for a given temperature $k_BT_0=0.7\Delta$ and fixing
$\gamma_1=0.5\Delta$, $\gamma_2=1.0\Delta$, $\epsilon_1=0.4\Delta$, $\epsilon_2=1.9\Delta$, $\Gamma=0.3$, $s_1=1$ and $s_2=-1$.
Note that $L_{12}=L_{21}$ only when $\delta\phi=n\pi$, with $n$ an integer.

\begin{figure}[!htb]
	\centering
	\includegraphics[width=0.9\columnwidth]{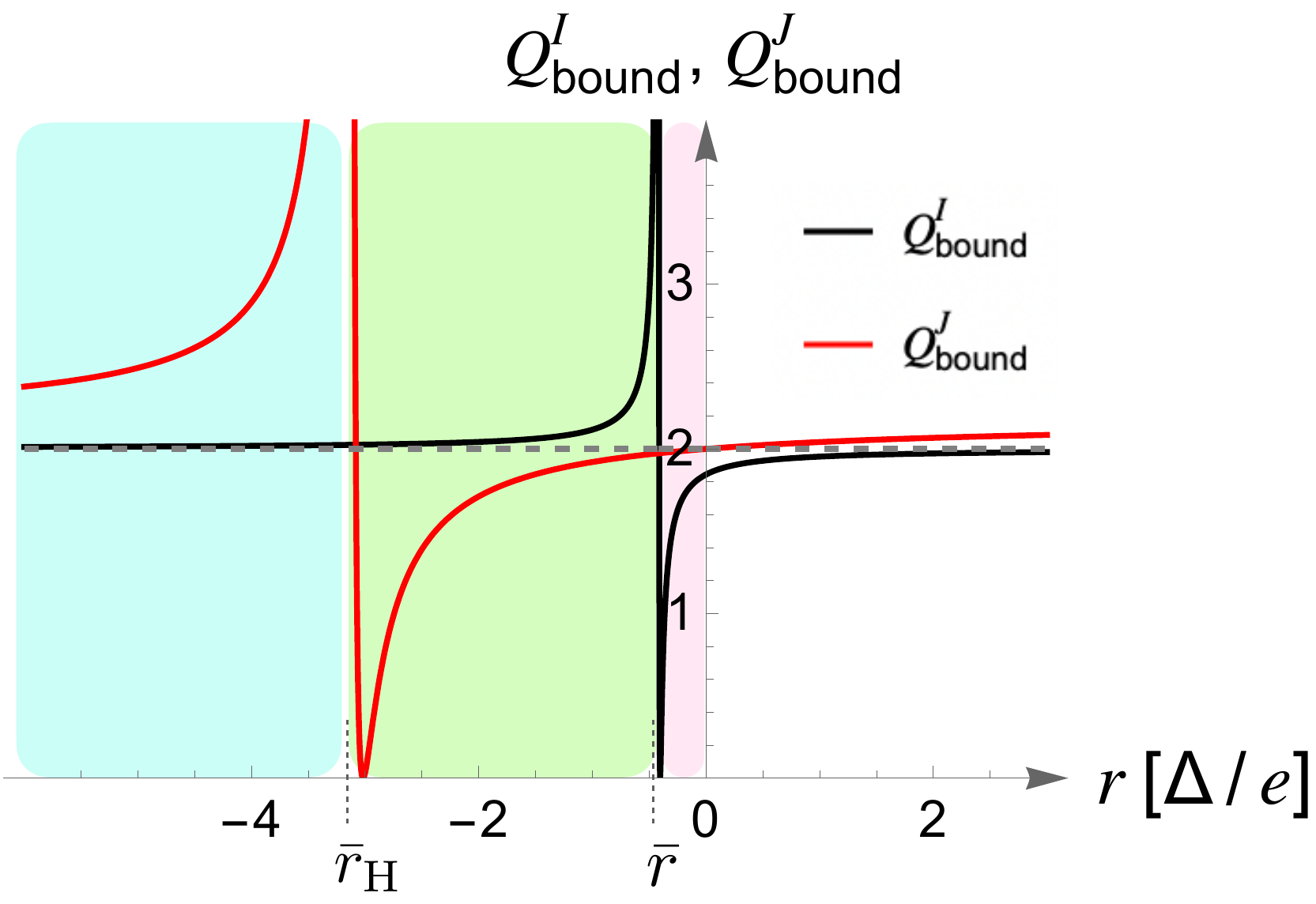}
	\caption{ $Q_{\rm bound}^I$ (black) and $Q^J_{\rm bound}$ (red) as functions of $r=VT_0/\Delta T$ in units of $\Delta/e$. In the region highlighted in pink (light blue) the Andreev interferometer works as a heat engine (refrigerator). In the intermediate region highlighted in light green the Andreev interferometer works as a heat pump. We used the same parameters of Fig.~\ref{phi}.}
	\label{qbound}
\end{figure}

For such a choice of parameters, in Fig.~\ref{qbound} we plot $Q_{\rm bound}^I$ (black) and $Q^J_{\rm bound}$ (red) as functions of the ratio $r=VT_0/\Delta T$ in units of $\Delta/e$.
We first note that by varying $r$ the two quantities $Q_{\rm bound}^I$ and $Q^J_{\rm bound}$ can take any value greater or equal to zero.
Differently from $Q_{\rm bound}^J$, $Q_{\rm bound}^I$ departs substantially from 2 only in a relatively small range of values of $r$. Moreover, they are both very close to 2 as long as $r>0$, i.e.~the voltage and the temperature difference are concordant.
In particular, since both $L_{12}$ and $L_{21}$ are positive for this parameters' choice, $Q_{\rm bound}^I$ and $Q^J_{\rm bound}$ reach zero for two different {\it negative} values of $r$.
Namely, $Q_{\rm bound}^I=0$ when $r=-(L_{12}+L_{21})/(2L_{11})$, while $Q^J_{\rm bound}=0$ when $r=-2L_{22}/(L_{12}+L_{21})$.
On the other hand, note that for $r=0$ we have $Q^J_{\rm bound}=2$ and  $Q_{\rm bound}^I=1/2\left(  1+L_{21}/L_{12}\right)^2$.

Furthermore, both bounds $Q_{\rm bound}^I$ and $Q^J_{\rm bound}$ can actually diverge.
For the first quantity, this happens at $r=\bar{r}=-L_{12}/L_{11}$ and for the second one at $r=\bar{r}^{\rm H}=-L_{22}/L_{21}$.
Interestingly, $\bar{r}$ corresponds to the ``stopping voltage''~\cite{Benenti2017}, which is the maximum voltage for which the system behaves as a heat engine, i.e.~produces an output power while absorbing heat.
Indeed, in the linear response the output power $P$ can be expressed in terms of the Onsager matrix elements as
\begin{equation}
P=-V\left(L_{11}\frac{V}{T_0}+L_{12}\frac{\Delta T}{T_0^2}
\right) ,
\end{equation}
so that $P$ is positive as long as $V$ is negative and
\begin{equation}
\label{he1}
V>-\frac{L_{12}}{L_{11}T_0}\Delta T\equiv V_{\rm stop}.
\end{equation}
The corresponding range of values of $r$ is highlighted in pink in Fig.~\ref{qbound}.
On the other hand, $\bar{r}^{\rm H}$ is related to the ``stopping temperature''~\cite{Benenti2017}, which is the maximum temperature difference, for a given applied voltage, for which the system works as a refrigerator, i.e.~extract heat from the cold electrode.
Fixing $\Delta T>0$, from Eq.~(\ref{Ons}) one can see that $J<0$ as long as $\Delta T<-VT_0L_{21}/L_{22}\equiv \Delta T_{\rm stop}$ (assuming that $L_{21}$ is positive), which can be written in terms of $r$ as $r<\bar{r}^{\rm H}$.
Such a range of values is highlighted in light blue in Fig.~\ref{qbound}.
Finally, for $r$ in the range $\bar{r}^{\rm H}<r<\bar{r}$, highlighted in green in Fig.~\ref{qbound}, the system behaves as a heat pump, i.e.~heat flows out of the hot electrode ($J>0$) while power is provided to the system ($P<0$).
Interestingly, $Q_{\rm bound}^I$ is strictly smaller than 2 (i.e.~the bound is looser with respect to the B-S bound) when the Andreev interferometer works as a heat engine, while $Q^J_{\rm bound}$ is strictly larger than 2 (i.e.~the bound is tighter with respect to the B-S bound) when it works as a refrigerator.
Such a correlation between the operating mode of the system and the specific departure of the bounds from the B-S value is due to a change in the behavior of the currents.
Indeed, $Q^I_{\rm bound}$ goes below 2 when the charge current crosses zero becoming positive, i.e.~the system starts operating as a heat engine.
Similarly, $Q^J_{\rm bound}$ goes above 2 when the heat current crosses zero becoming negative, i.e.~the system starts operating as a refrigerator.
Finally, the asymptotic values (for $r\rightarrow \pm\infty$) of the two bounds are $Q_{\rm bound}^I\sim 2$ and $Q_{\rm bound}^J\sim \frac{1}{2}\left(  1+\frac{L_{12}}{L_{21}}\right)^2$.

\section{Conclusions}
\label{SecConc}
In this  paper we have derived new bounds for TURs in the absence of TRS for steady-state electronic systems in a two-terminal setup.
Such bounds  ($Q_{\rm bound}^I$ and $Q_{\rm bound}^J$) constrain the value of 
the two quantities $Q^I=S_I\sigma/(I^2k_B)$ and $Q^J=S_J\sigma/(J^2k_B)$, which are associated to charge ($I$) and heat ($J$) currents, respectively.
Here $S_I$ ($S_J$) is the charge (heat) current noise, $\sigma$ is the entropy production rate and $k_B$ is the Boltzmann constant.
While in systems which preserve TRS the two bounds are equal to 2 ($Q_{\rm bound}^I=Q_{\rm bound}^J=2$, Barato-Seifert bound~\cite{Barato2015}),
we have found that, in the absence of TRS, the two bounds are different and depend both on the details of the system and on the ratio between the affinities (applied voltage and temperature difference).
In particular, they differs from 2 as long as the off-diagonal Onsager matrix coefficients ($L_{12}$ and $L_{21}$) are different, and can take any positive value.
Moreover, we have found that $Q_{\rm bound}^I=2$ when $\Delta T=0$ and that $Q^J_{\rm bound}=2$ when $V=0$.

We have then focused on the case of coherent conductors, where the scattering approach can be used.
Notably, for a normal conductor attached to two terminals one finds, even when TRS is broken, that $L_{12}= L_{21}$, so that both bounds equal 2.
We have then calculated the two bounds for the case of a hybrid superconducting system, where, in general, $L_{12}\neq L_{21}$.
In
the specific case of an Andreev interferometer (TRS is broken when a superconducting phase difference exists between two superconducting regions), we have investigated the behavior of $Q_{\rm bound}^I$ and $Q_{\rm bound}^J$
as functions of ratio between affinities.
Remarkably, we have found that $Q_{\rm bound}^I$ is always smaller than 2 when the Andreev interferometer works as a heat engine, while $Q^J_{\rm bound}$ is always larger than 2 when the Andreev interferometer works as a refrigerator.

We conclude by noting that the occurrence of TRS breaking in materials, especially in unconventional superconductors\cite{Saykin2022}, is an important information which is not easy to obtain.
Remarkably, the new bounds we have found can be used to detect the absence of TRS in superconducting materials.
On the other hand, an interesting development of the present paper would be the study of these new bounds for TURs in the absence of TRS for non-superconducting systems, such as a normal conductor which is partially coherent or where transport is predominantly classical, in which cases $L_{12}$ can be different from $L_{21}$. For example, a system composed by quantum dots in the Coulomb blockade regime, or a system where decoherence is partially introduced by a fictitious dephasing terminal.

\section{Acknowledgements}
R.F. acknowledges financial support from PNRR MUR project PE0000023-NQSTI.
F.T. acknowledges the Royal Society through the International Exchanges between the UK and Italy (Grants No. IEC/R2/212041).

\begin{appendix}

\section{Currents in a normal multi-channel system}
\label{AppNormal}
For a normal system, charge $I$ and heat $J$ currents flowing in the right lead can be calculated, using the scattering approach~\cite{Blanter2000}, as
\begin{align}
\label{I}
I=\frac{e}{h} \int_{-\infty}^{\infty}dE\; 
D(E) [f_R(E)-f_L(E)]
\end{align}
and
\begin{align}
\label{J}
J=\frac{1}{h} \int_{-\infty}^{\infty}dE\; 
D(E) (E-eV) [f_R(E)-f_L(E)] ,
\end{align}
where $D(E)$ is the transmission probability from left to right and $f_L$ and $f_R$ are the Fermi distribution functions in the left and right reservoir, respectively.
The right reservoir is kept at a voltage $V$ and temperature $T_0+\Delta T$, while the left reservoir is grounded and at a temperature $T_0$.

Expanding $f_L$ and $f_R$ in linear response, one finds
\begin{align}
L_{11}=\frac{e^2T_0}{h} \int_{-\infty}^{\infty}dE\; 
D(E) [-f_0'(E)] ,
\end{align}
\begin{align}
L_{22}=\frac{T_0}{h} \int_{-\infty}^{\infty}dE\; 
D(E) E^2 [-f_0'(E)] ,
\end{align}
\begin{align}
L_{12}=\frac{eT_0}{h} \int_{-\infty}^{\infty}dE\; 
D(E) E [-f_0'(E)]
\end{align}
and $L_{21}=L_{12}$, even in the absence of TRS.

\section{Currents in a hybrid superconducting multi-channel system}
\label{AppSuper}
Within the scattering approach~\cite{Claughton1996,Lambert1998}, the charge $I$ and heat $J$ currents flowing in a hybrid superconducting system can be calculated at the N right lead and can be written, respectively, as
\begin{align}
I= & \frac{e}{h}\sum_{j={\rm L,R}}\sum_{\alpha=\pm 1}(\alpha)\int_0^{\infty} dE\;\left[ \delta_{{\rm R}j}N_{\rm R}^\alpha (E) f_{\rm R}^\alpha(E)-  \right. \nonumber\\ &\left. \sum_{\beta=\pm 1} P_{{\rm R}j}^{\alpha\beta}(E) f_j^\beta(E) \right] ,
\end{align}
and
\begin{align}
J=&\frac{1}{h}\sum_{j={\rm L,R}}\sum_{\alpha=\pm 1}\int_0^{\infty} dE\;(E-eV)\left[ \delta_{{\rm R}j}N_{\rm R}^\alpha (E) f_{\rm R}^\alpha(E)-\right. \nonumber\\ &\left. \sum_{\beta=\pm 1} P_{{\rm R}j}^{\alpha\beta}(E) f_j^\beta(E) \right] ,
\end{align}
where $\alpha,\beta=\pm 1$ stand for particles/holes-like quasiparticles, $f_i^\alpha(E)$ is the Fermi distribution function in lead $i={\rm L,R}$ for particle-like quasiparticles ($\alpha=1$) or hole-like quasiparticles ($\alpha=-1$),
$P_{ij}^{\alpha\beta}(E)$ is the probability, at energy $E$, for a $\beta$-like quasiparticle in lead $j$ to be scattered into a $\alpha$-like quasiparticle in lead $i$ and $N^\alpha_i(E)$ is the number of open channels, at energy $E$, in lead $i$ for $\alpha$-like quasiparticle.
In terms of the scattering matrix elements, the probabilities can be written as~\cite{Lambert1998}
\begin{align}
P_{ij}^{\alpha\beta}(E)=\sum_{a,b}\left|
s_{(i,a),(j,b)}^{\alpha,\alpha}(E)
\right|^2 ,
\end{align}
where $s_{(i,a),(j,b)}^{\alpha,\beta}(E)$ is the scattering amplitude for a $\beta$-like quasiparticle arriving from channel $b$ in lead $j$ at energy $E$ to be scattered as a $\alpha$-like quasiparticle into lead $i$, channel $a$, with $a=1,\dots,N_i^\alpha(E)$ and $b=1,\dots,N_j^\beta(E)$.
One can prove that if TRS is preserved we have
\begin{align*}
P_{ij}^{\alpha\beta}(E)=P_{ji}^{\beta\alpha}(E).
\end{align*}
Note that Eqs.~(\ref{I}) and (\ref{J}) can be obtained from the above equations  by setting the Andreev scattering coefficients to zero.

Here we use the following notations: $R^{\alpha\alpha}=P_{\rm RR}^{\alpha\alpha}$, $R_a^{\alpha\beta}=P_{\rm RR}^{\alpha\beta}$ (with $\alpha\ne\beta$), $T^{\alpha\alpha}=P_{\rm LR}^{\alpha\alpha}$,
$T'^{\alpha\alpha}=P_{\rm RL}^{\alpha\alpha}$, $T_a^{\alpha\beta}=P_{\rm LR}^{\alpha\beta}$ (with $\alpha\ne\beta$), $T_a'^{\alpha\beta}=P_{\rm RL}^{\alpha\beta}$ (with $\alpha\ne\beta$).
Using the fact that
\begin{align}
N_i^\alpha(E)=\sum_{\beta,j}P^{\alpha\beta}_{ij} (E),
\end{align}
$L_{12}$ can be written as
\begin{align}
\label{L12a}
L_{12}=&\frac{eT_0}{h} \int_{0}^{\infty}dE\; E[T'^{++}(E)+T_a'^{+-}(E)-\nonumber \\
&T'^{--}(E)-T_a'^{-+}(E)] \left( -\frac{\partial f}{\partial E}\right) ,
\end{align}
while using 
\begin{align}
N_j^\beta(E)=\sum_{\alpha,i}P^{\alpha\beta}_{ij} ,
\end{align}
$L_{21}$ can be written as
\begin{align}
\label{L21a}
L_{21}=&\frac{eT_0}{h} \int_{0}^{\infty}dE\; E[T^{++}(E)+T_a^{-+}(E)-\nonumber \\
&T^{--}(E)-T_a^{+-}(E)] \left( -\frac{\partial f}{\partial E}\right)  .
\end{align}
Moreover, 
\begin{align}
\label{L22a}
L_{22}=&\frac{T_0}{h} \int_{0}^{\infty}dE\; E^2[T'^{++}(E)+T_a'^{+-}(E)+\nonumber \\
&T'^{--}(E)+T_a'^{-+}(E)] \left( -\frac{\partial f}{\partial E}\right) .
\end{align}
Note that in a SN system and at low temperature ($k_BT_0\ll\Delta$) the above equations yield $L_{12}=L_{21}=L_{22}\simeq 0$, since all transmissions are zero for energies below the gap $\Delta$.

\section{Composition of the scattering matrix of an Andreev interferometer}
\label{AppScattering}

In order to obtain the scattering matrix of the Andreev interferometer depicted in Fig.~\ref{AI}, we compose~\cite{Datta1997} the scattering matrices for the individual components.
The three-leg beam splitter is described by the following, energy-independent, scattering matrix~\cite{Buttiker1984}
\begin{align}
S_{\rm B}=
\begin{pmatrix}
-s_1 \sqrt{1-2\Gamma} & \sqrt{\Gamma}  & \sqrt{\Gamma} \\
\sqrt{\Gamma}  & a & b \\
\sqrt{\Gamma}  & b & a
\end{pmatrix} ,
\end{align}
where $a=1/2(s_2+s_1\sqrt{1-2\Gamma})$, $b=1/2(-s_2+s_1\sqrt{1-2\Gamma})$, $0\leq\Gamma\leq 1/2$, $s_{1,2}=\pm$.
The two resonant barriers, labelled by 1 and 2, are described by the following reflection and transmission amplitudes
\begin{align}
r_{\rm b, 1(2)}(E)=(E-\epsilon_{1(2)})\tau_{1(2)} (E)
\end{align}
and
\begin{align}
t_{\rm b, 1(2)}(E)=-i\gamma_{1(2)}\tau_{1(2)} (E) ,
\end{align}
respectively, where
\begin{align}
\tau_{1(2)} (E) =\frac{1}{(E-\epsilon_{1(2)})+i\gamma_{1(2)}} .
\end{align}
Here $\gamma_{1(2)}$ is the coupling strength of each individual barrier of the resonator 1(2) and $\epsilon_{1(2)}$ is the position of the resonant level for the barrier 1(2).
Each ideal NS interface, labelled by 1 and 2, is described by the following perfect Andreev reflection amplitude
\begin{align}
r^{\rm he}_{\rm A,1(2)} (E)= \exp (-i {\rm Arccos} \left ( \frac{E}{|\Delta_{1(2)}|} \right) -i\phi_{1(2)}),
\end{align}
for an electron to be reflected as a hole, and
\begin{align}
r^{\rm eh}_{\rm A,1(2)} (E)= \exp (-i {\rm Arccos} \left ( \frac{E}{|\Delta_{1(2)}|} \right) +i\phi_{1(2)}),
\end{align}
for a hole to be reflected as an electron, where $\Delta_{1(2)}$ is the superconducting order parameter and $\phi_{1(2)}$ is the superconducting phase.

\end{appendix}

\bibliography{TUR}

\end{document}